\documentclass[aps,twocolumn,groupedaddress]{revtex4}

\usepackage{epsfig}
\usepackage{graphicx}
\usepackage[T1]{fontenc}
\usepackage{ae}
\usepackage{color}
\usepackage[latin1]{inputenc}

\newcommand{\di}{\unitlength0.5cm\begin{picture}(1,1)
\put(0.,0.2){\circle*{0.25}} \put(0.6,0.2){\circle*{0.25}}
\put(0,0.2){\line(1,0){0.6}}
\end{picture}}

\newcommand{\dii}{\unitlength0.6cm\begin{picture}(1,1)
\put(0.,0.2){\circle*{0.25}} \put(0.6,0.2){\circle*{0.25}}
\put(0,0.12){\line(1,0){0.6}} \put(0,0.29){\line(1,0){0.6}}
\end{picture}}

\newcommand{\tv}{\unitlength0.6cm\begin{picture}(1,1.5)
\put(0.,0.2){\circle*{0.25}} \put(0.6,0.2){\circle*{0.25}}
\put(1.2,0.2){\circle*{0.25}} \put(0,0.2){\line(1,0){1.2}}
\end{picture}}

\begin{document}


\title[HTE for Heisenberg models]
{Tenth-order high-temperature
expansion for the
susceptibility and the specific heat of spin-$s$ Heisenberg models with
arbitrary
exchange patterns: Application to pyrochlore and kagome magnets}

\author{Andre Lohmann$^1$, Heinz-J\"urgen Schmidt$^2$
and Johannes Richter$^1$}
\email{johannes.richter@physik.uni-magdeburg.de}

\affiliation{$^1$Institut f\"ur Theoretische Physik, Otto-von-Guericke-Universit\"at Magdeburg,\\
PF 4120, D - 39016 Magdeburg, Germany\\
$^2$Universit\"at Osnabr\"uck, Fachbereich Physik,
Barbarastr. 7, D - 49069 Osnabr\"uck, Germany}

\begin{abstract}
We present the high-temperature expansion (HTE) up to 10th order of the specific heat
$C$ and the
uniform susceptibility $\chi$ for Heisenberg models with arbitrary
exchange patterns and arbitrary spin quantum number $s$. We encode the
algorithm in a C++ program provided in the  supplementary material and available at {\tt
http://www.uni-magdeburg.de/jschulen/HTE10/} which allows to get
explicitly the HTE
series for concrete Heisenberg models.
We apply our algorithm to pyrochlore ferromagnets and kagome
antiferromagnets using several  Pad\'e approximants for the HTE series.
For the pyrochlore ferromagnet we use the HTE data for $\chi$ to estimate the Curie
temperature $T_c$ as a function of the  spin quantum number $s$. We find that
$T_c$ is smaller than that for the simple cubic lattice, although both
lattices have the same coordination number.
For the kagome antiferromagnet the influence of the spin quantum number $s$
on the susceptibility as a function of renormalized temperature $T/s(s+1)$ is rather
weak for temperatures down to $T/s(s+1) \sim 0.3$. On the other hand, the
specific heat as a function of $T/s(s+1)$ noticeably depends on $s$. The
characteristic maximum in $C(T)$ is monotonously shifted to lower values of
$T/s(s+1)$ when increasing $s$.
\end{abstract}

\maketitle

\section{Introduction}\label{sec:I} Magnetic systems described by the
Heisenberg Hamiltonian
\begin{equation}\label{model} H=\sum_{\mu <\nu}
J_{\mu \nu} {\bf s}_\mu \cdot {\bf s}_\nu
\end{equation}
are an active field of theoretical and experimental
research.\cite{lnp} The accurate description of these quantum
many-body systems is a basic aim of theoretical investigations. The
comparison with experimental studies typically requires the
calculation of the temperature dependence of physical properties,
such as the susceptibility $\chi$ and the specific heat $C$. For
unfrustrated spin systems  the quantum Monte Carlo technique is a
suitable tool to provide precise data, but it is not applicable due
to the sign problem for frustrated quantum spin models.\cite{TrWi05}
Hence, reliable results for strongly frustrated
quantum spin models are notoriously rare. Since there is a very
active research in the field of frustrated quantum magnetism, see
e.g. Refs.~\onlinecite{lnp,balents,Nisoli} and references therein,
theoretical methods to calculate thermodynamic quantities in the
presence of frustration are highly desirable. One of the most
interesting systems is the kagome Heisenberg antiferromagnet which
we will consider in Sec.~\ref{kagome}. This highly frustrated
magnetic system has been extensively investigated theoretically for
spin quantum number $s=1/2$, see, e.g.,
Refs.~\onlinecite{Leung1993,Waldtmann1998,Mambrini2000,Singh2007,Sindzingre2009,Evenbly2010,
Yan2011,lauchli2011,sakai2011,goetze,depenbrock,normand2013,becca2013},
 and in the classical limit
$s \to \infty$, see, e.g.,
 Refs.~\onlinecite{chalker1992,reimers1993,huber01,Zhito08,Moessner13}.
On the experimental side, several kagome compounds have $s>1/2$, however,
 the theoretical study of quantum models  with $s>1/2$ lags behind.

A  universal straightforward approach to calculate thermodynamic
quantities for unfrustrated as well as  frustrated
magnetic systems is the high-temperature expansion (HTE). For
Heisenberg models this method was introduced  in an early work by
W.~ Opechowski\cite{opechowski} based on a method of approximate
evaluation of the partition function developed by H.A. Kramers. In
the 1950ies and 1960ies the method was further developed and widely
applied to various Heisenberg systems, see e.g.
Refs.~\onlinecite{wood1955,wood1957,wood1958,wood1967,dalton1969}.

The HTE method is now well-established and its application to magnetic
systems is a basic tool in theoretical physics, see
Refs.~\onlinecite{domb_green,OHZ06} and references therein. For the
Heisenberg model with nearest-neighbor interaction  on standard
lattices now typically the HTE is known up to high-orders, see for
example Refs.~\onlinecite{elstner1993}, \onlinecite{elstner1994}, and
\onlinecite{singh2012}
where the HTE up to 14th order for the triangular and up to
16th order for the kagome and for the hyperkagome lattices were published.
On the other hand, often
magnetic compounds and corresponding spin models are of interest,
where two, three or even more exchange constants are relevant.
Typical examples are magnetic systems with nearest-neighbor,
next-nearest-neighbor, and 3rd-nearest-neighbor couplings, see,
e.g., Ref.~\onlinecite{reuther2011}. Moreover, in most of the
quasi-low-dimensional magnetic compounds interchain or interlayer
couplings play a role, see, e.g.
Ref.~\onlinecite{inter_chain}. Note further that available high-order HTE
are often restricted to spin quantum number $s=1/2$,  see, e.g.,
Refs.~\onlinecite{elstner1993}, \onlinecite{elstner1994},
\onlinecite{singh2012},
\onlinecite{oitmaa1995} and
\onlinecite{kapella} as an
example. As a rule, for such more complex exchange geometries and/or
higher spin quantum numbers $s>1/2$  relevant for the interpretation
of experimental data the HTE is not available in higher orders. An
earlier attempt to bridge this gap was published in
Ref.~\onlinecite{SSL01}, where general analytical HTE expressions
were given for arbitrary $s$ and arbitrary Heisenberg exchange
couplings  up to order three. Very recently the present authors have
published  a significant extension of this work using computer
algebraic tools.\cite{HTE_1} In that paper the HTE algorithm  for
general spin-$s$ Heisenberg models up to 8th order was presented. This
algorithm was encoded  as a C++ program. The download (URL {\tt
http://www.uni-magdeburg.de/jschulen/HTE/}) and use are free. Thus
this algorithm provides a flexible tool for the community to compute
the HTE for the susceptibility and the specific heat, which can be
used to analyze the thermodynamics of spin models, to check
approximations, and, last but not least, to compare experimental
data with model predictions.

In our previous work\cite{HTE_1} we considered several models. Our
results demonstrated that the 8th order HTE with
a subsequent  Pad\'e approximation is (i) able to describe correctly the
maximum of the susceptibility of a square-lattice s=1/2 Heisenberg
antiferromagnet, (ii) can yield better results for spin systems in dimension $d>1$
than full exact diagonalization, and (iii) gives good agreement with Monte Carlo
data for the classical pyrochlore antiferromagnet down to temperatures of
about 40\% of the exchange coupling.

The aim of the present paper is twofold. On the one hand, we
will extend our earlier approach\cite{HTE_1} up to 10th order for
general Heisenberg Hamiltonians  (Sec II). Again we provide this new
extended tool as a freely accessible C++ program, see
supplementary material\cite{supp} and also {\tt
http://www.uni-magdeburg.de/jschulen/HTE10/}, were except the code 
a manual on how to use the code can also be found.
We recommend to use {\tt http://www.uni-magdeburg.de/jschulen/HTE10/}, since
there will be provided available updates of the code (fixing of bugs,
improvement of the code, etc.).
On the other hand, results for the
susceptibility and the specific heat for the spin-$s$ Heisenberg
model on the pyrochlore and the kagome lattices are provided (Sec
III), which we will use to discuss the influence of the spin quantum
number on the thermodynamics of these models.  In the Appendices
\ref{app_pyro} and \ref{app_kagome} we present explicitly the HTE
series for the pyrochlore and the kagome spin-$s$ Heisenberg
magnets. In the supplementary part\cite{supp} we present the detailed
results for the 10th order HTE of specific heat and susceptibility
and the auxiliary quantities to be introduced in Sect \ref{sec:E}.

\section{Brief explanation of the method}\label{sec:E}
We consider the HTE expansion of extensive quantities $f$, e.~g.,
susceptibility  $\chi$ or specific heat $C$, of the form
\begin{equation}\label{E1}
f^{\Sigma}(\beta)=\sum_{n=0}^\infty c_n^{\Sigma,f}\,\beta^n \;.
\end{equation}
Here $\beta$ is the inverse dimensionless temperature
$\beta=\frac{J}{k_B T}$, where $J$ is a typical exchange energy, and
the index $\Sigma$ indicates the dependence on the spin system
$\Sigma$ which is given by the Hamiltonian (\ref{model}) and the
value of the spin quantum number $s$. A further dependence on the
magnetic field is possible but neglected in this paper. As mentioned
in the Introduction we do not consider special systems $\Sigma$
but rather look for a general HTE expansion valid for arbitrary Heisenberg systems.

Interestingly, the coefficients $ c_n^{\Sigma,f}$ in (\ref{E1}) can
be written in the form of scalar products between two vectors $Q$
and $p$ such that the first vector $Q$ only depends on the spin
system $\Sigma$, but not on $s$, and the second one $p$ only on the
considered quantity $f$ and the spin quantum number $s$. The index
set of the vectors $p$ and $Q$ can be identified with finite sets
$G_n^f$ of multigraphs, see Ref.~\onlinecite{HTE_1} for the details. Thus the
scalar product of $Q$ and $p$ is a sum over multigraphs
${\mathcal G}\in G_n^f$:
\begin{equation}\label{E2}
 c_n^{\Sigma,f}=\sum_{{\mathcal G}\in G_n^{f}}Q^{\Sigma}({\mathcal G})\,p^f({\mathcal G})
\;.
\end{equation}
To give an elementary example, consider $f=\chi$, the zero-field uniform
susceptibility. Simplifying the notation a bit we may write
\begin{eqnarray}\nonumber
\chi(\beta)&=& Q({\mathcal G}_0)\,p_0\,\beta+ Q({\mathcal G}_1)\,p_1\,\beta^2+\\
\label{E3} && \left(Q({\mathcal G}_2)\,p_2 + Q({\mathcal
G}_3)\,p_3\right)\,\beta^3+ {\mathcal O}(\beta^4) \;.
\end{eqnarray}
Here
\begin{eqnarray}\label{E4a}
p_0&=&\frac{1}{3}r,\quad r\equiv s(s+1)\\
\label{E4b}
p_1&=&-\frac{2}{9}r^2,\;p_2=-\frac{1}{18}r^2,\;p_3=\frac{2}{27}r^3,\\
\label{E4c} {\mathcal
G}_0&=&{\footnotesize\bullet}\;,\;{\mathcal G}_1=\di,\;
{\mathcal G}_2=\dii,\;
{\mathcal G}_3=\tv\\
\label{E4d}
Q({\mathcal G}_0)&=&N\quad \mbox{(number of spins)},\\
\label{E4e} Q({\mathcal G}_1)&=&\sum_{\mu<\nu}J_{\mu\nu},\;
Q({\mathcal G}_2)=\sum_{\mu<\nu}J_{\mu\nu}^2,\\
\label{E4f} Q({\mathcal
G}_3)&=&\sum_{\lambda<\mu<\nu}J_{\lambda\mu}J_{\mu\nu} \;.
\end{eqnarray}
In this example the $p^f({\mathcal G})$ are polynomials in the
variable $r\equiv s(s+1)$ of the form $p^f({\mathcal
G})=\sum_{\nu=0}^n a_\nu\,r^\nu$, where $n$ is the order of HTE in
(\ref{E2}). This holds in general. Also generally the
$Q^{\Sigma}({\mathcal G})$ are polynomials in the coupling constants
$J_{\mu\nu}$ that can be calculated by considering the various ways
of embedding the graph ${\mathcal G}$ into the spin system. For
example, each mapping of the $3$-chain $\;\tv\;$ onto three spins
with numbers $\lambda<\mu<\nu$ gives rise to a term
$J_{\lambda\mu}J_{\mu\nu}$ in $Q^{\Sigma}({\mathcal G})$. The
condition $\lambda<\mu<\nu$ guarantees that different embeddings
resulting from symmetries of ${\mathcal G}$ are counted only once.
Also this is typical for the general situation.

If the spin system is an infinite lattice, the $Q^{\Sigma}({\mathcal
G})$ have to be redefined by first considering finite realizations
of $\Sigma$, and then dividing by the number of spins $N$ and
considering the thermodynamic limit $N\longrightarrow\infty$. If
${\mathcal G}$ is not connected then $Q^{\Sigma}({\mathcal G})$
would scale with $N^c$, where $c$ is the number of connected
components of ${\mathcal G}$. Hence, for sake of consistency, the
sets $G_n^f$ must consist of connected graphs only. Keeping this in
mind, the coefficients $c_n^{\Sigma,f}$ obtained represent rigorous
results on infinite spin lattices that are notoriously rare.

For the determination of $Q^{\Sigma}({\mathcal G})$ there exist
effective computer programs. On the other hand, we have determined
the ``universal" polynomials $p^f({\mathcal G})$ for $f=\chi,C$ and
$\gamma({\mathcal G})\le 10$, see Ref.~\onlinecite{supp}. The method used has
been explained to some details in Ref.~\onlinecite{HTE_1} and will only be
sketched here. The crucial auxiliary data are the polynomials
$p^{(t)}({\mathcal G})(r)$ resulting from the moments of the
Heisenberg Hamiltonian (\ref{model}) via
\begin{equation}\label{E5}
\tilde{t}_n\equiv\mbox{Tr}\,H^n = (2s+1)^n \sum_{{\mathcal G}\in
G_n^{(t)}}Q^{\Sigma}({\mathcal G})\,p^{(t)}({\mathcal G}) \;.
\end{equation}
In this case the polynomials can be shown to be of the form
$p^{(t)}({\mathcal G})=\sum_{\nu=g}^\gamma a_\nu\,r^\nu$, where
$g=g({\mathcal G})$ is the number of vertices of ${\mathcal G}$ and
$\gamma=\gamma({\mathcal G})$  the number of edges. These
polynomials have been essentially determined by numerically
calculating $\mbox{Tr}\,H^n$ and $Q^{\Sigma}({\mathcal G})$ for a
suitable number of randomly chosen spin systems and then solving the
linear system of equations (\ref{E5}) for the $p^{(t)}({\mathcal
G})$. This has to be repeated for different values of
$s=1/2,\,1,\,3/2,\ldots$ in order to estimate the rational
coefficients of the polynomials $p^{(t)}({\mathcal
G})(r),\;r=s(s+1)$. Additionally, partial analytical results from
Refs.~\onlinecite{domb_green} and \onlinecite{HTE_1} have been used and
various cross checks have been performed.

As described in Ref.~\onlinecite{HTE_1}, one deduces from the
$p^{(t)}({\mathcal G})$  the ``magnetic moments"
\begin{equation}\label{E6}
\tilde{\mu}_n\equiv\mbox{Tr}\left(S_3^2\,H^n\right) = (2s+1)^n
\sum_{{\mathcal G}\in G_n^{(m)}}Q^{\Sigma}({\mathcal
G})\,p^{(m)}({\mathcal G}) \;,
\end{equation}
and from the $\tilde{t}_n$ and $\tilde{\mu}_n$ the HTE series for
$C(\beta)$ and $\chi(\beta)$. In this last step there will occur
products of $Q^{\Sigma}({\mathcal G}_\mu)$ as well as contributions
from disconnected graphs. In order to obtain manifestly extensive
quantities, these various non-extensive terms have to cancel by
means of certain ``product rules" of the form
\begin{equation}\label{E7}
Q^{\Sigma}({\mathcal G}_\mu)\,Q^{\Sigma}({\mathcal
G}_\nu)=\sum_\lambda\,c_{\mu\nu}^\lambda\,Q^{\Sigma}({\mathcal
G}_\lambda) \;.
\end{equation}
All calculations described in this section involve a total number of
$7355$ graphs. Especially those steps leading from the moments to
the HTE series have been performed with the aid of the computer
algebra system MATHEMATICA 8.0.

\section{Applications}
\label{applic}

The region of validity of the HTE can be extended by  Pad\'e
approximants\cite{baker61} (see also
Refs.~\onlinecite{domb_green} and \onlinecite{OHZ06}). The Pad\'e
approximants are ratios of two
polynomials $[m,n]=P_m(x)/R_n(x)$ of degree $m$ and $n$ and they provide an
analytic continuation of a function $f(x)$ given by a power series.
As a rule approximants with $m \sim n$ provide best results.
Since we have a power series up to 10th order, we use here the corresponding
[4,6], [5,5], and [6,4]  Pad\'e approximants.
As in our previous paper\cite{HTE_1} we will present the temperature
dependence of physical quantities  using a renormalized temperature $T/s(s+1)$.

\begin{figure}[ht]
\begin{center}
\includegraphics[clip=on,width=80mm,angle=0]{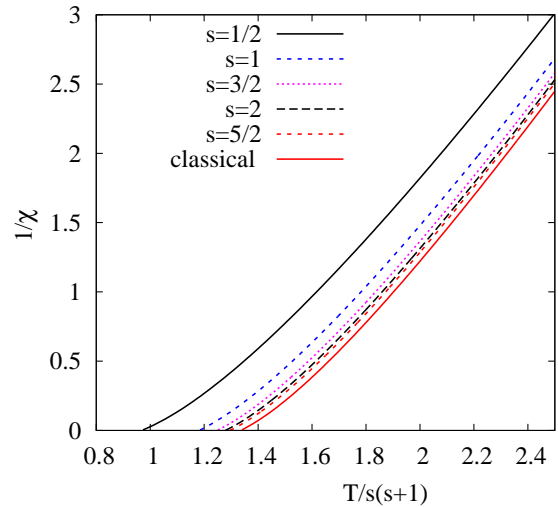}
\end{center}
\caption{(Color online)  Pad\'e approximant [4,6] of the inverse susceptibility $1/\chi$ of
the
pyrochlore Heisenberg
ferromagnet for various spin quantum numbers
$s$.
\label{inv_chi}}
\end{figure}

\begin{figure}[ht]
\begin{center}
\includegraphics[clip=on,width=80mm,angle=0]{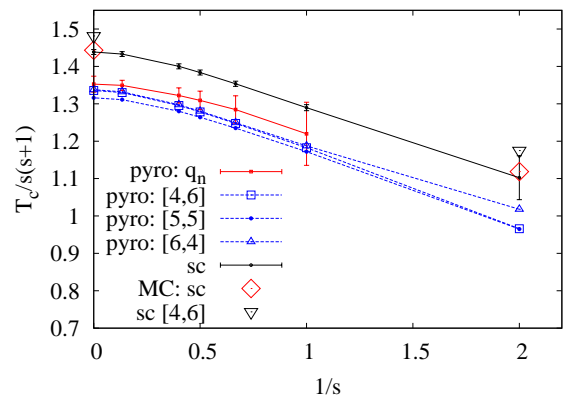}
\end{center}
\caption{(Color online) Curie temperature $T_c$ in dependence on the inverse spin quantum number
$1/s$
of the pyrochlore Heisenberg
ferromagnet for $s=1/2$, $1$, $3/2$, $2$, $5/2$, $15/2$ and $s \to \infty$.
For comparison we show also the  $T_c$ values for the simple-cubic
Heisenberg ferromagnet.
The Monte Carlo
 data (MC) for $s=1/2$ and $s \to \infty$ are taken from
Ref.~\onlinecite{QMC_s12} and from Ref.~\onlinecite{MC_clas},
respectively.
\label{curie}}
\end{figure}

\subsection{The pyrochlore Heisenberg ferromagnet}

The pyrochlore antiferromagnet  has attracted much attention over
the last years as an example of a  highly frustrated  three-dimensional (3D) magnetic
system, see, e.g. Refs.~\onlinecite{moessner01,bramwell01,monopol}
and references therein. In our previous paper\cite{HTE_1} we have
already presented the analytical expressions for $\chi$ and $C$ up
to order 8 as well as the temperature dependence of the
susceptibility for the pyrochlore Heisenberg antiferromagnet, where
a very good agreement of the [4,4] Pad\'e approximant
  with classical Monte Carlo results down to temperatures of
about 40\% of the exchange coupling was found.
The new terms of HTE in orders 9 and 10 can be found in
Appendix~\ref{app_pyro}.

Here we consider the ferromagnetic case. We consider only nearest
neighbor bonds and set $J_{\mu \nu}=J=-1$ for neighboring sites
$\mu$ and  $\nu$. We want to demonstrate that the 10th order HTE is
an appropriate tool to determine the critical (Curie) temperature
$T_c$ for 3D ferromagnets. To the best of our knowledge so far no
data for $T_c$ of the pyrochlore ferromagnet were reported in the
literature. In Fig.~\ref{inv_chi} we show the  Pad\'e approximant
[4,6] of the inverse susceptibility $1/\chi$. We see the typical
behavior of 3D ferromagnet. The extreme quantum case $s=1/2$ is
somewhat separated from the other curves, but for $s>1$ the curves
are very close to each other. The zeros of $1/\chi(T)$ curves can be
understood as an estimate of the critical temperature. More
sophisticated methods exploit the behavior of the expansion
coefficients $c_n$, see Eq.~(\ref{E1}), of the susceptibility to
determine $T_c$, see, e.g.
Refs.~\onlinecite{wood1955,wood1958,yeomans}. One variant is to
analyze the quotient $q_n=c_n/c_{n-1}$. If the critical  behavior of
$\chi$ is given by $\chi(T) \propto (T-T_c)^{-\lambda}$, $T \to
T_c+0$, in the limit $n\to \infty$ this quotient depends linearly on
$1/n$ according to
$q_n= \frac{kT_c}{J} + (\lambda-1)\frac{kT_c}{J n}$.
Hence we get $\frac{kT_c}{J}$ by $\lim_{n\to \infty} q_n =\frac{kT_c}{J}$.

We made a linear fit of our HTE data
for $q_n$ including data points for $n=5,\ldots,10$ to get an
approximate value for $T_c$. Our results for $T_c$ are shown in
Fig.~\ref{curie}. For comparison we also show $T_c$ data for the
simple-cubic ferromagnet, where precise Monte Carlo data are
available for $s=1/2$ (Ref.~\onlinecite{QMC_s12}) and for $s \to \infty$
(Ref.~\onlinecite{MC_clas}), which yield an impression of the accuracy of the HTE
estimate of $T_c$. Except for the $s=1/2$-pyrochlore case  the  $q_n$
data follow reasonably well a straight line (see also the error bars
in Fig.~\ref{curie}). For the $s=1/2$-pyrochlore ferromagnet the linear
fit of the  $q_n$ data due to extremely large fluctuations of the
data fails.\cite{shift} The comparison with the Monte Carlo data for
the simple-cubic ferromagnet demonstrates that indeed the HTE series
up to order 10 for the susceptibility may yield accurate values for
$T_c$. Already the poles in the Pad\'e approximants provide
reasonable results (we have about 14\% deviation from Monte Carlo
data for $s=1/2$ and about 9\% for $s \to \infty$). The linear fit
of $q_n$ is even very close to the Monte Carlo results.

Comparing the pyrochlore and simple cubic lattices  we find that $T_c$
is significantly
lower for the pyrochlore lattice.
(Note that a simple molecular field approximation would lead to
identical values of $T_c$, since both lattices have the same
coordination number.)
A similar finding was reported in Ref.~\onlinecite{stacked},
 where the Curie temperatures of stacked  square and a stacked kagome
 ferromagnets were compared.
In analogy to the discussion in Ref.~\onlinecite{stacked} we may
attribute the lower $T_c$ values of the pyrochlore lattice to
geometric frustration. For the ferromagnetic ground state
frustration is irrelevant, i.e. the  ground-state energies are
identical for the pyrochlore and simple-cubic ferromagnets. However,
due to frustration the upper bound of the eigenenergies (related to
the absolute value of ground-state energy of the corresponding
antiferromagnet) is much lower  for the pyrochlore ferromagnet than
that for the simple-cubic lattice. Hence, one can expect that
excited states with antiferromagnetic spin correlations have lower
energy for the pyrochlore ferromagnet resulting in a larger
contribution to the partition function at a certain finite
temperature $T$ in comparison with the simple-cubic ferromagnet.

\begin{figure}[ht]
\begin{center}
\includegraphics[clip=on,width=80mm,angle=0]{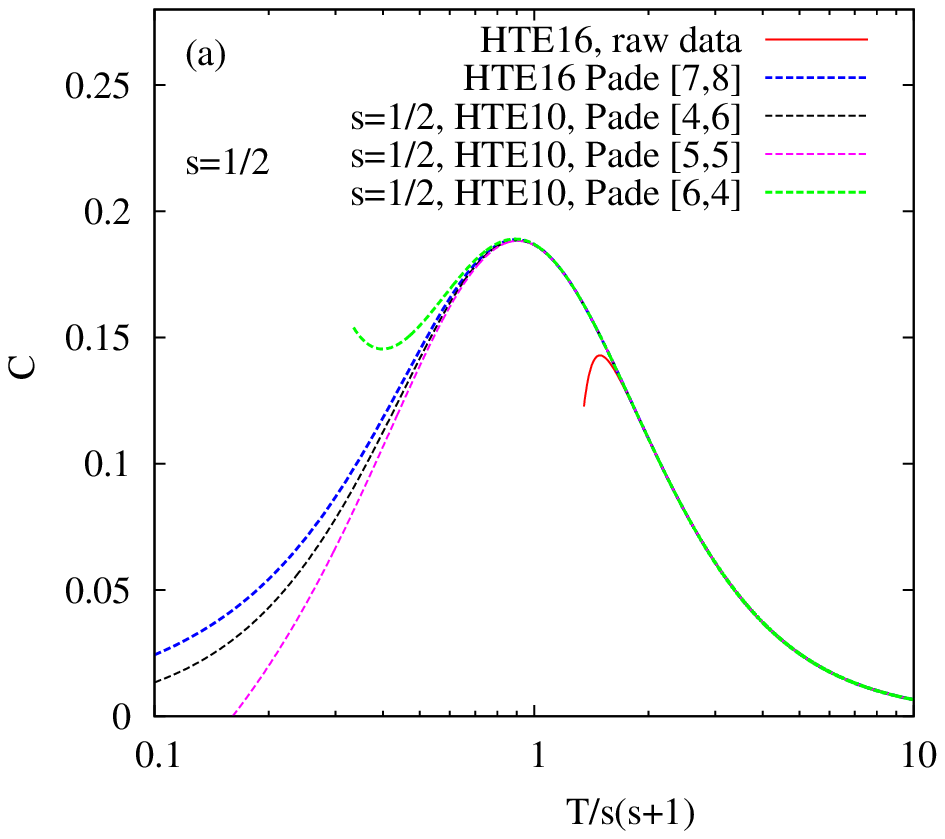}
\includegraphics[clip=on,width=80mm,angle=0]{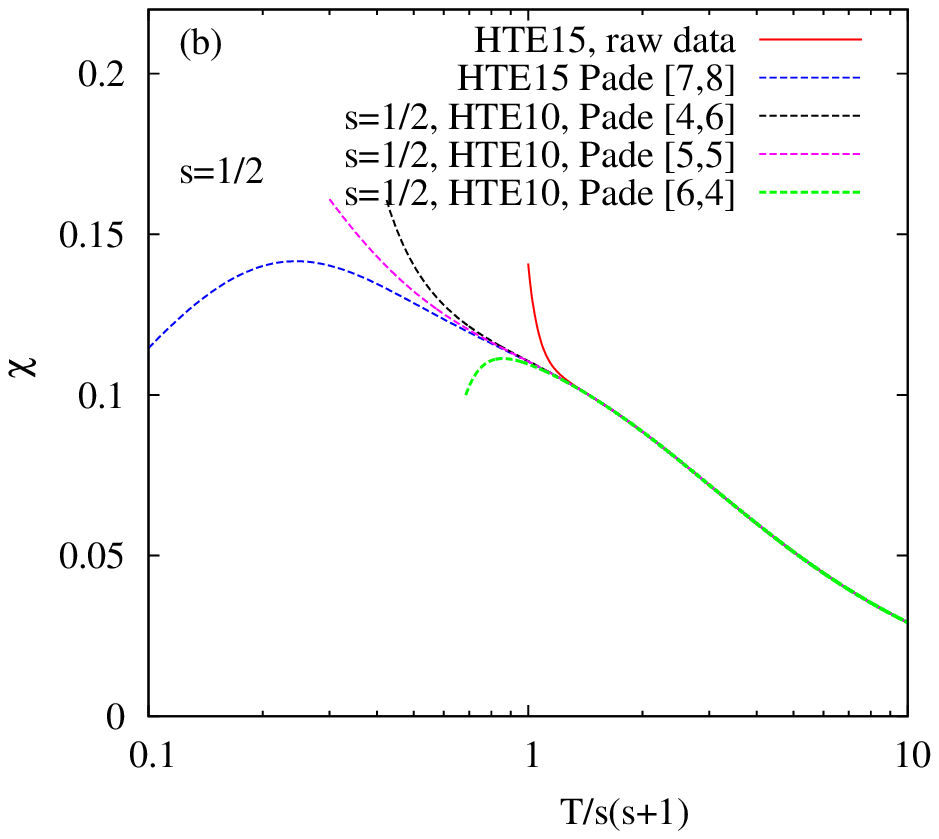}
\end{center}
\caption{(Color online) Specific heat $C$ (a) and susceptibility $\chi$ (b) of the
$s=1/2$ kagome Heisenberg antiferromagnet. For comparison we show
the raw data of the 15th/16th order HTE and  the  corresponding
Pad\'e [7,8] approximant taken from Ref.~\onlinecite{elstner1994}.
\label{kago_s12}}
\end{figure}

\begin{figure}[ht]
\begin{center}
\includegraphics[clip=on,width=80mm,angle=0]{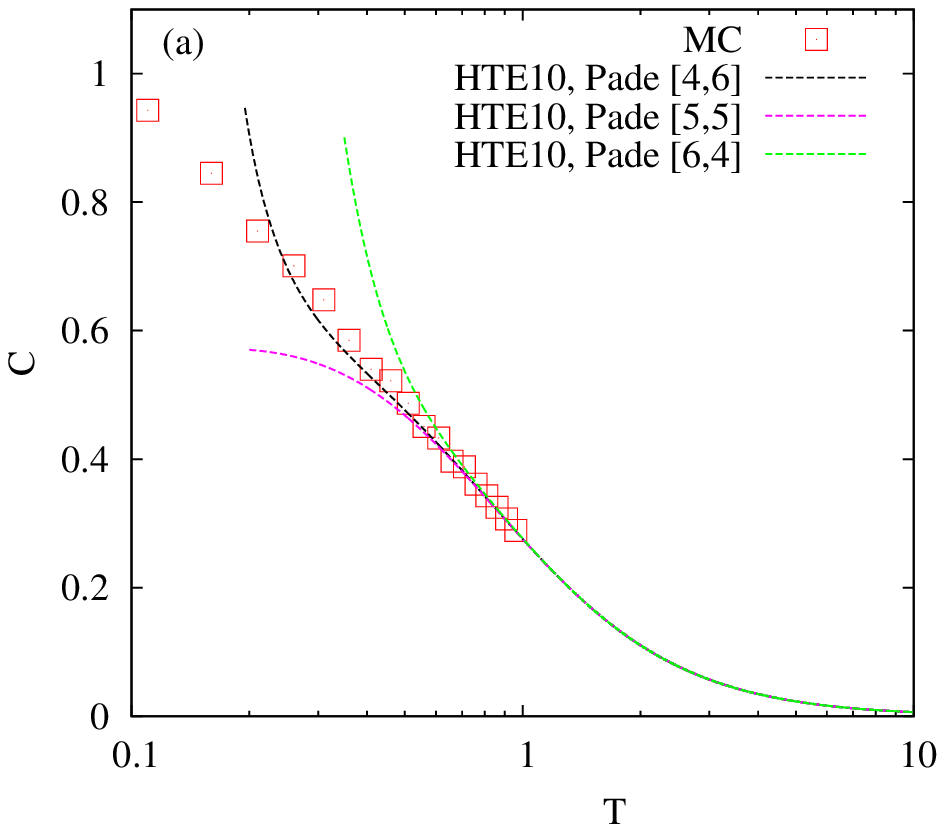}
\includegraphics[clip=on,width=80mm,angle=0]{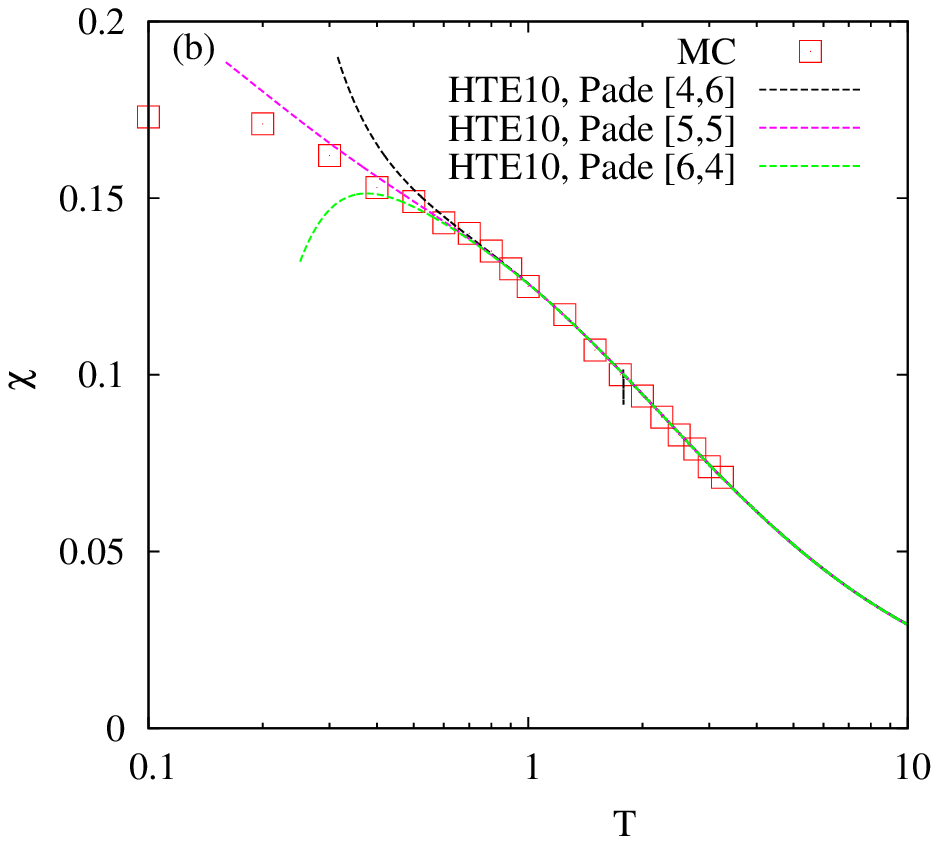}
\end{center}
\caption{(Color online) Specific heat $C$ (a) and susceptibility $\chi$ (b) of the
classical kagome Heisenberg antiferromagnet. For comparison we show
the Monte Carlo data (MC)
 taken from Ref.~\onlinecite{huber01}.
\label{kago_clas}}
\end{figure}

\subsection{The kagome Heisenberg antiferromagnet}
\label{kagome}
The two-dimensional (2D) kagome antiferromagnet is one of the most interesting and challenging spin
models. There are
numerous papers investigating the ground state of the $s=1/2$ case, see,
e.g.,
Refs.~\onlinecite{harris1992,sachdev1992,Leung1993,Waldtmann1998,Mambrini2000,Singh2007,Sindzingre2009,Evenbly2010,
Yan2011,lauchli2011,sakai2011,goetze,depenbrock,normand2013,becca2013}
 and references therein, but so far
no conclusive answer on the nature of
the ground state and the existence of a spin gap has been found.
The finite-temperature properties are also widely discussed for the
spin-$1/2$ model, including the analysis of HTE
series.\cite{elstner1994,nakamura1995,tomczak1996,sindz2000,canals2002,misguich2005,rigol2007,misguich2007}
On the other hand, there are several kagome compounds with spin quantum number
$s>1/2$. We mention, the jarosite compounds with $s=5/2$ (see, e.g.,
Refs.~\onlinecite{jaros1,jaros2}), the magnetic compounds
KCr$_3$(OD)$_6$(SO$_4$)$_2$ (Ref.~\onlinecite{s32_a}) and
SrCr$_{9p}$Ga$_{12-9p}$O$_{19}$ (Ref.~\onlinecite{s32_b})  with $s=3/2$, and  the recently
synthesized  BaNi$3$(OH)$_2$(VO$4$)$_2$ (Ref.~\onlinecite{s1})  compound with $s=1$.
For the $s=5/2$ one may expect that  a classical
Monte Carlo
approach\cite{chalker1992,reimers1993,huber01,Zhito08,Moessner13} might be reasonable, but
for $s=1$ and for $s=3/2$ certainly quantum effects are important.
However, we will see that at least for the specific heat the classical
Monte Carlo data significantly deviate from the data for $s=5/2$, see below.

\begin{figure}
\begin{center}
\includegraphics[clip=on,width=80mm,angle=0]{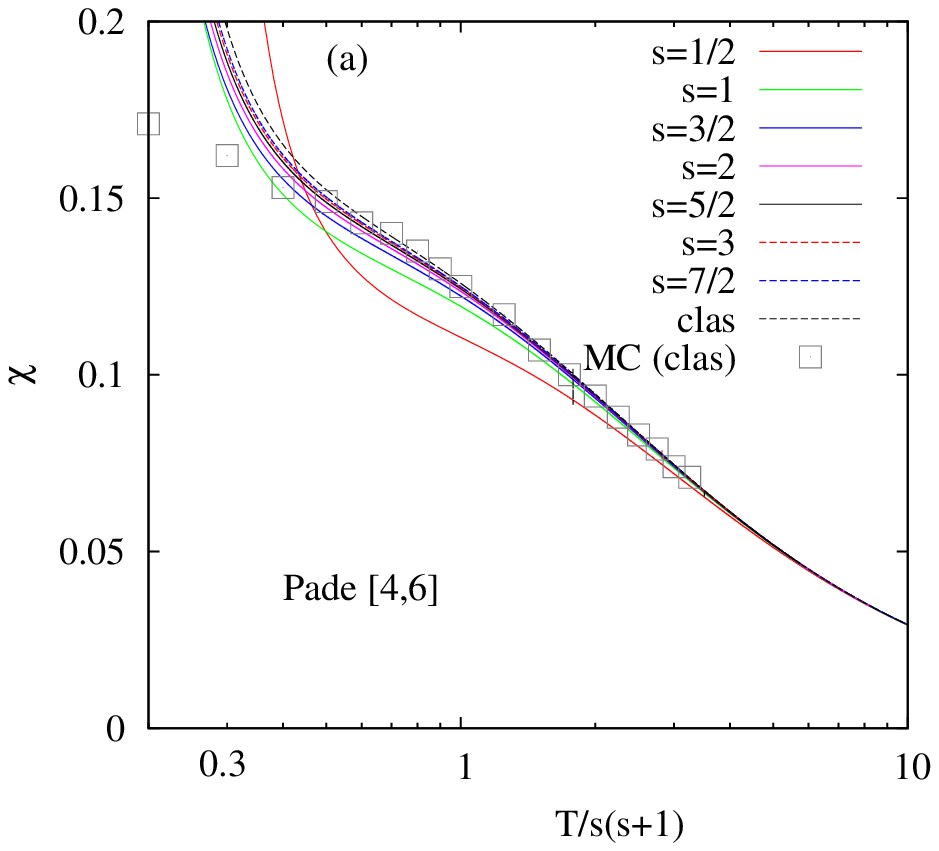}
\includegraphics[clip=on,width=80mm,angle=0]{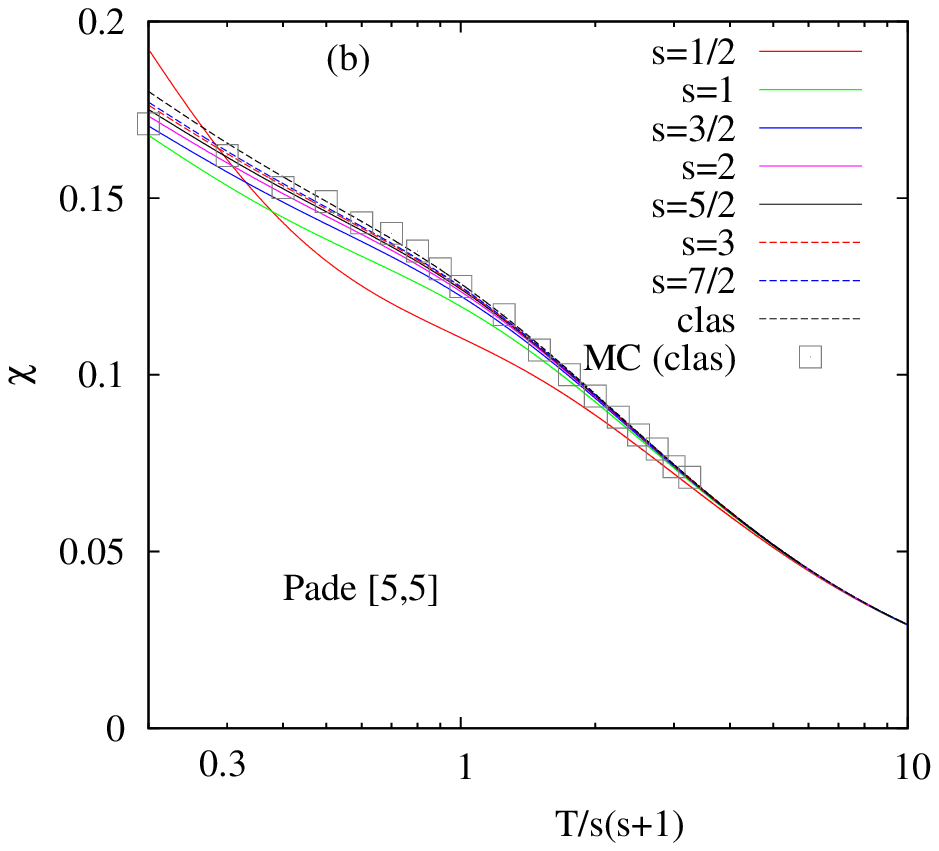}
\end{center}
\caption{(Color online) HTE data for the susceptibility $\chi$
of the
spin-$s$ kagome Heisenberg
antiferromagnet, (a)  Pad\'e [4,6], (b)  Pad\'e [5,5].
 For comparison we show the Monte Carlo data (MC)
 taken from Ref.~\onlinecite{huber01}.
\label{kago_chi}}
\end{figure}
\begin{figure}
\begin{center}
\includegraphics[clip=on,width=80mm,angle=0]{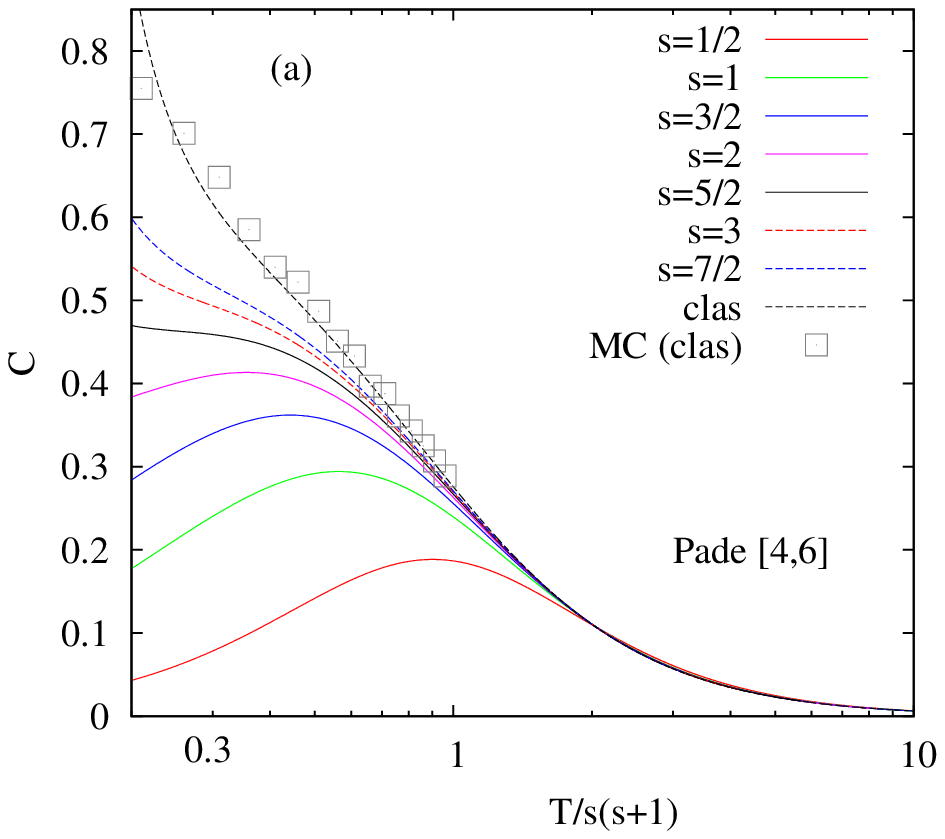}
\includegraphics[clip=on,width=80mm,angle=0]{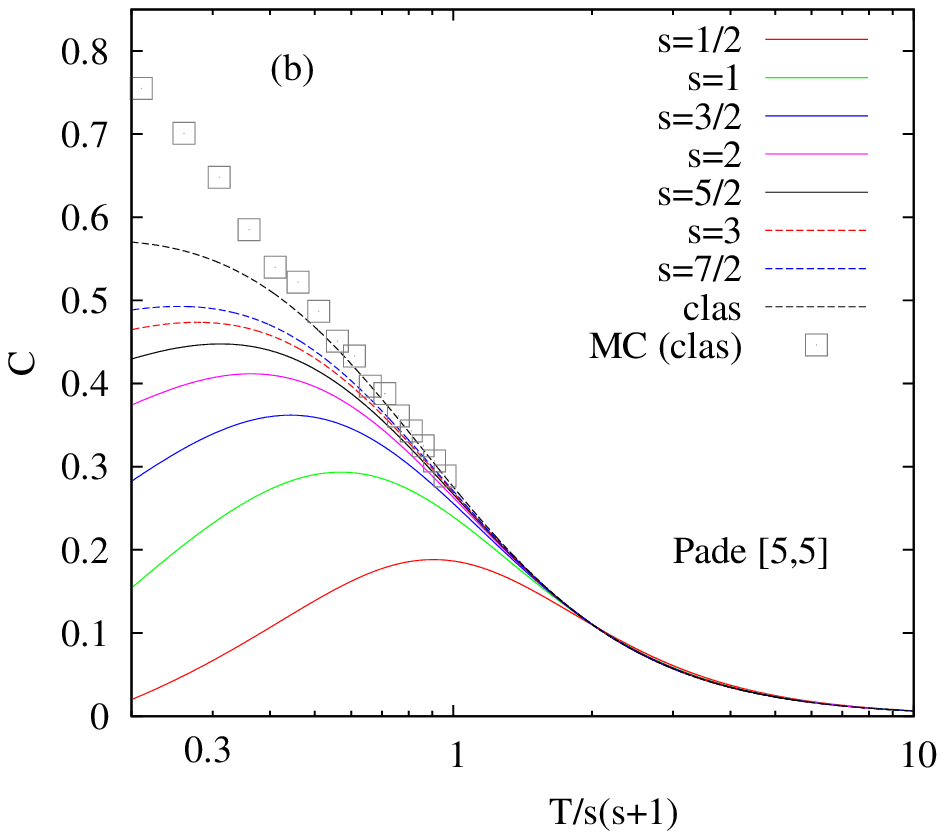}
\end{center}
\caption{(Color online) Specific heat $C$
of the
spin-$s$ kagome Heisenberg
antiferromagnet,  (a)  Pad\'e [4,6], (b)  Pad\'e [5,5].  For comparison we show the Monte Carlo data
(MC)
 taken from Ref.~\onlinecite{huber01}.
\label{kago_C}}
\end{figure}
\begin{figure}
\begin{center}
\includegraphics[clip=on,width=80mm,angle=0]{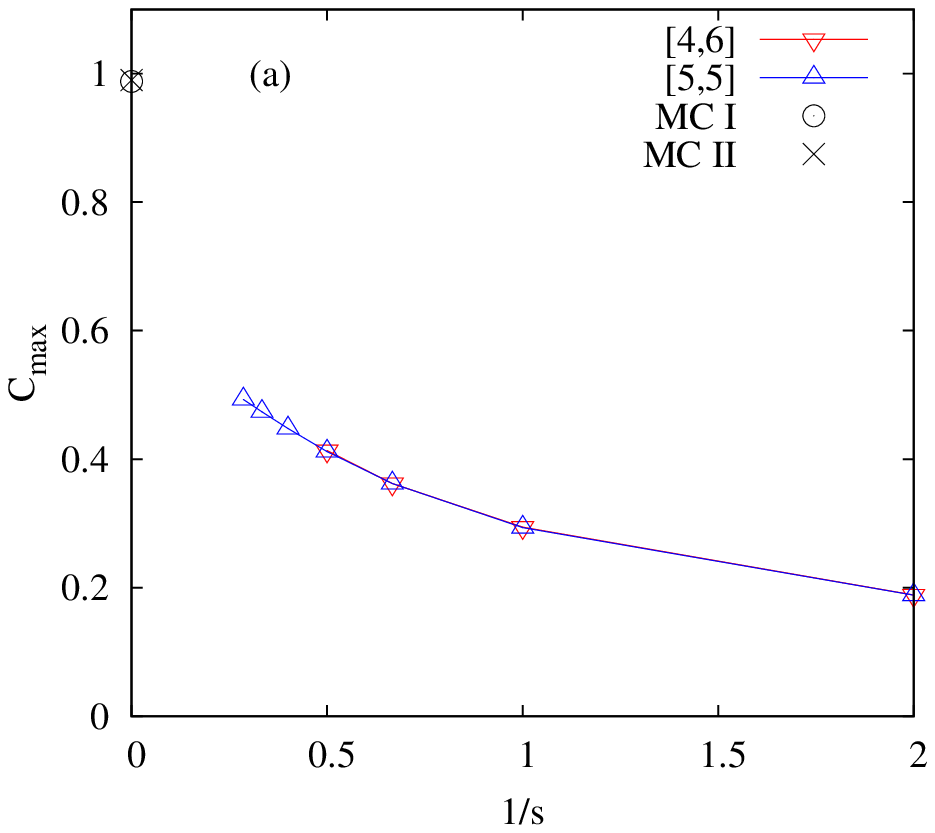}
\includegraphics[clip=on,width=80mm,angle=0]{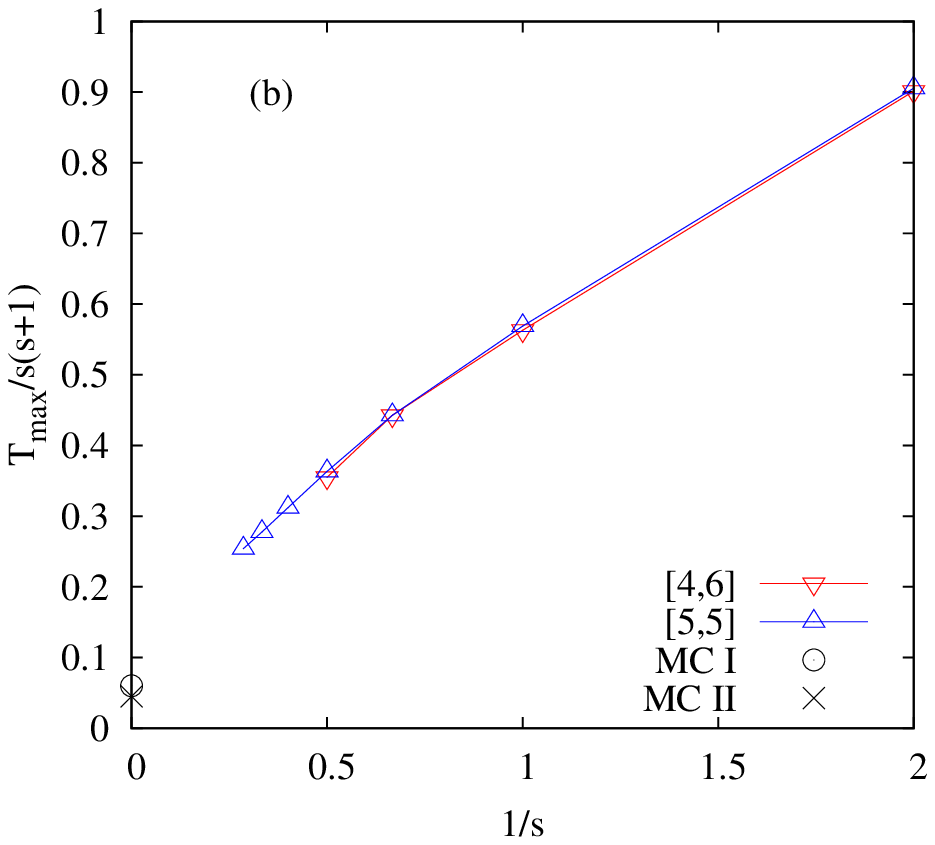}
\end{center}
\caption{(Color online) Height $C_{max}$ (a) and position $T_{max}$ (b) of the  maximum in the specific heat $C$  in dependence on the inverse spin
quantum
number $1/s$
of the kagome Heisenberg
antiferromagnet.  The Monte Carlo results for $s \to \infty$
 are taken from Ref.~\onlinecite{huber01} (MC I)
 and Ref.~\onlinecite{Zhito08} (MC II). Note, however, that the maximum
 in $C(T)$ for the classical model is not well-pronounced, rather there is
a fairly broad region of high values of $C$.\cite{Zhito08}
\label{max_kago}}
\end{figure}

We present the HTE series for $\chi$ and $C$ up to order 10 and for
arbitrary $s$ in Appendix~\ref{app_kagome}. Remember that in
Ref.~\onlinecite{elstner1994}  the HTE series for $s=1/2$ for $\chi$
($C$) was given up to order 15 (16).\cite{fehler} As a benchmark
test we first compare in Figs.~\ref{kago_s12} and \ref{kago_clas}
our HTE-Pad\'e data with available data for $s=1/2$
model\cite{elstner1994} and for the classical model.\cite{huber01}
This comparison leads to the conclusion that (i) the [4,6] and [5,5]
Pad\'e approximants are favorable and that (ii) our HTE-Pad\'e data
are quite accurate down to temperatures $T/s(s+1) \sim 0.5 $ ($T
\sim 0.4$) for $s=1/2$ ($s \to \infty$). In particular, the maximum
in $C$ present for the $s=1/2$ model at  $T/s(s+1) \sim 0.9$, cf.
Refs.~\onlinecite{elstner1994,nakamura1995,tomczak1996,sindz2000,canals2002,misguich2005,rigol2007},
is correctly described by our  Pad\'e approximants. Note, however,
that for $s=1/2$  there are indications for a second low-temperature
maximum in $C(T)$ below $T/s(s+1) = 0.1$, see
Refs.~\onlinecite{elstner1994,tomczak1996,sindz2000,misguich2005},
which is not covered by our HTE approach. Another  characteristic
features is the shoulder present in $\chi(T)$ for $s=1/2$ at about
$T/s(s+1) = 1$, which is also well described by our Pad\'e
approximants.

In Figs.~\ref{kago_chi} and \ref{kago_C} we compare the $\chi(T)$
(Fig.~\ref{kago_chi}) and $C(T)$ data (Fig.~\ref{kago_C}) for spin
quantum numbers $s=1/2,1,\ldots,7/2,\infty$. The susceptibility data
clearly show that all curves for $s>1/2$ form a narrow bundle  in
the temperature range accessible by our approach. Only for $s=1/2$
the  $\chi(T)$ curve is out of this bundle. Hence, one can argue
that for $s>1/2$ quantum effects in $\chi$ are almost negligible at
normalized temperatures $T/s(s+1) \gtrsim 0.4$. The situation is
quite different for the specific heat $C$, cf.~Fig.~\ref{kago_C}.
The maximum in $C(T)$, already mentioned above for $s=1/2$, is
evidently dependent on the spin quantum number $s$: Its position
$T_{max}/s(s+1)$ is shifted to lower normalized temperatures and
its  height $C_{max}$ increases with growing $s$. Hence, the quantum
effects seem to be important also for quite large values of $s$. The
basic difference in the influence of $s$ on $\chi(T)$ and $C(T)$ can
be attributed to an exceptional density of low-lying singlet
excitations, see e.g. Ref.~\onlinecite{Waldtmann1998}. These
nonmagnetic excitations are irrelevant for $\chi$ but important for
$C$. Hence, our HTE-Pad\'e data for $C(T)$ can be understood as an
indirect indication for the existence of a unusual large density of
low-lying singlet excitation also for $s>1/2$. We show the position
$T_{max}/s(s+1)$ and the  height $C_{max}$ as a function of $1/s$
in  Fig.~\ref{max_kago}. From Fig.~\ref{kago_C}a it is obvious that for the
[4,6]-Pad\'e approximant a maximum is exists only for $s<5/2$.
The tendency of how the classical limit is approached is clearly
visible from our HTE data. There is indeed a remarkably  strong dependence on the spin
quantum number. The slope of the corresponding curves shown in
Fig.~\ref{max_kago} is even increasing when approaching the
classical limit $1/s = 0$.
Let us mention that for the above discussion of the maximum of
the specific heat the 8th order HTE presented in our previous
paper\cite{HTE_1} is not
appropriate, since the corresponding [4,4] Pad\'e approximant exhibits an
unphysical pole in the vicinity of the maximum for
$s > 1/2$.

\section{Conclusions}

In this paper we present the HTE series  up to 10th order
of the specific heat $C$
and the uniform susceptibility $\chi$ for Heisenberg models with arbitrary exchange
patterns $J_{\mu \nu}$ and spin quantum
number $s$. Our HTE scheme is encoded in a C++ program using as input the
exchange
matrix  $J_{\mu \nu}$ and  spin quantum
number $s$. Using Pad\'e approximants for the HTE series the scheme can be used to
discuss thermodynamic
properties of general Heisenberg systems down to moderate temperatures of
about $T/s(s+1) \sim 0.4 \ldots 0.5 $ and thus for the interpretation
of experimental data in rather wide temperature  range, especially if other precise methods such as the quantum
Monte Carlo method or the finite-temperature  density matrix renormalization
group approach are not applicable.
We apply our scheme to the 3D pyrochlore ferromagnet to calculate  the
Curie
temperature $T_c$ in dependence on the  spin quantum number $s$. Comparing
$T_c$ of the pyrochlore ferromagnet with corresponding values for the simple-cubic
ferromagnet we
find that the triangular configuration of bonds present in the pyrochlore
lattices leads to a noticeable lowering
of $T_c$.
Using our HTE scheme for the kagome antiferromagnet we discuss the influence
of $s$ on the temperature dependence of $C$ and $\chi$. While the effect of
$s$ on $\chi$ in the accessible temperature range is rather weak, there is a
well-pronounced
shift of the maximum in the temperature dependence of specific heat to
lower renormalized temperatures $T/s(s+1)$ when increasing $s$.
\\

\vspace{1cm}

{\bf Acknowledgment}\\
JR thanks  A. Hauser and M. Maksymenko for fruitful discussions.

\appendix

\section{The high-temperature
expansion for the susceptibility and the specific heat for the
Heisenberg model on the pyrochlore lattice}
\label{app_pyro}

The general formulas for the  susceptibility and the specific heat
for the Heisenberg model on the pyrochlore lattice with NN exchange
constant $J$ up to 8th order can be found in
Ref.~\onlinecite{HTE_1}. For the sake of consistency with this reference
we have set  $\beta=\frac{1}{k_B T}$ in appendix A and B which is slightly different
from the definition in section \ref{sec:E}.
 The formulas for the  9th and
10h order read for the susceptibility
\begin{widetext}
\begin{eqnarray}  \label{HTE_chi_pyro}
&&\chi(\beta)=\frac{N}{J}\sum_{n=1}^\infty c_n (J \beta )^n\\
&&c_9=\frac{1}{5143824000}r^2(-2710665+142840908r-2195288001r^2+14497581366r^3\nonumber\\
&& \; \; -45972407664r^4+77794619872r^5-82650432896r^6+46730617088r^7)\nonumber\\
&&c_{10}=-\frac{1}{169746192000}r^2(51519240-2994073848r+51386055291r^2\nonumber\\
&& \; \; -396940170060r^3+1579391570694r^4-3442568263344r^5+4692701814464r^6\nonumber\\
&& \; \; -4374573206272r^7+2124654831616r^8).\nonumber
\end{eqnarray}
and for the specific heat 
\begin{eqnarray}  \label{HTE_C_pyro}
&&C(\beta)=N k \sum_{n=2}^\infty d_n(J\beta)^n\\
&& d_9 = -\frac{1}{285768000} r^2 (-1807110 + 91861560 r - 1255862151 r^2 + 6268644864
r^3\nonumber\\
&& \; \;- 8882615472 r^4 - 1691186688 r^5 - 21317760 r^6 + 1042017280
r^7)\nonumber\\
&&d_{10}=-\frac{1}{6286896000}r^2(-25759620+1451298330r-22610800701r^2\nonumber\\
&& \; \; +142189820847r^3-349296723134r^4+154955752848r^5\nonumber\\
&& \; \; +102919717624r^6+82927576960r^7-11100907520 r^8),\nonumber
\end{eqnarray}
\end{widetext}
where $r=s(s+1)$.

\section{The high-temperature
expansion for the susceptibility and the specific heat for the
Heisenberg model on the kagome lattice}
\label{app_kagome}

The general formulas for the
Heisenberg
model on the kagome lattice with NN exchange constant $J$
read for the susceptibility
\begin{widetext}
\begin{eqnarray}  \label{HTE_chi_kago}
&&\chi(\beta)=\frac{N}{J}\sum_{n=1}^\infty c_n (J \beta )^n\\
&&c_1 = \frac{1}{3}r\nonumber \\
&&c_2 = -\frac{4}{9}r^2\nonumber \\
&&c_3 = \frac{1}{9}r^2 (-1 + 4 r)\nonumber \\
&&c_4 = -\frac{4}{405}r^2 (3 - 28 r + 37 r^2) \nonumber \\
&&c_5 = \frac{1}{4860}r^2 (-45 + 702 r - 1892 r^2 + 1328 r^3)\nonumber \\
&&c_6 = -\frac{1}{510300}r^2 (1728 - 35946 r + 164289 r^2 - 207896 r^3 + 102576 r^4)\nonumber \\
&&c_7 = \frac{1}{6123600}r^2 (-8694 + 218916 r - 1401381 r^2 + 2888772 r^3 -
2251248 r^4 + 909184 r^5)\nonumber \\
&&c_8 = -\frac{1}{22963500}r^2 (15390 - 446256 r + 3538764 r^2 - 10535337 r^3 +
12202552 r^4 - 7318640 r^5 + 2416640 r^6)\nonumber \\
&&c_9 = \frac{1}{7715736000}r^2 (-2710665 + 87954822 r - 807482331 r^2 +
3091042674 r^3 - 5118502560 r^4 + 4009481184 r^5\nonumber \\
&&- 2113197952 r^6 +
518354176 r^7)\nonumber \\
&&c_{10} = -\frac{1}{1273096440000} r^2 (257596200 - 9180862110 r + 93799827171
r^2 - 426255134022 r^3 + 931126345494 r^4\nonumber\\
&& - 977085756168 r^5 +
621427831616 r^6 - 280517703040 r^7 + 48779713280 r^8)\nonumber
\end{eqnarray}
and for the specific heat
\begin{eqnarray}  \label{HTE_C_kago}
&&C(\beta)=N k \sum_{n=2}^\infty d_n(J\beta)^n\\
&&d_2 = \frac{2}{3}r^2\nonumber\\
&&d_3 = -\frac{1}{9}r^2 (-3 + 4 r)\nonumber\\
&&d_4 = -\frac{2}{45}r^2 (-3 + 23 r + 3 r^2)\nonumber\\
&&d_5 = \frac{1}{162}r^2 (9 - 126 r + 116 r^2 + 48 r^3)\nonumber\\
&&d_6 = -\frac{1}{68040}r^2 (-1728 + 33426 r - 102969 r^2 - 19464 r^3 + 2144 r^4)\nonumber\\
&&d_7 = -\frac{1}{97200}r^2 (-1242 + 29556 r - 150039 r^2 + 96676 r^3 + 64544
r^4 + 20992 r^5)\nonumber\\
&&d_8 = \frac{1}{1093500}r^2 (7695 - 213084 r + 1435806 r^2 - 2537523 r^3 -
539132 r^4 + 58400 r^5 + 186680 r^6)\nonumber\\
&& d_9 = \frac{1}{214326000} r^2 (903555 - 28196370 r + 227949579 r^2 -
   634514526 r^3 + 285950568 r^4 + 230120832 r^5\nonumber\\
&& + 135526080 r^6 +   14890240 r^7)\nonumber\\
&&d_{10} = -\frac{1}{18860688000}r^2 (-51519240 + 1775187630 r - 16326321219 r^2 +
59250202038 r^3 - 69170925596 r^4\nonumber\\
&&- 15707506528 r^5 - 728311984 r^6 +
9196378240 r^7 + 3884989440 r^8).\nonumber
\end{eqnarray}
\end{widetext}


\end{document}